\documentclass{article}
\usepackage{alife10}
\usepackage{amsfonts}
\usepackage{amsmath}
\usepackage{graphicx}

\begin{document}

\title{The Role of Redundancy in the Robustness of Random Boolean Networks}
\author{Carlos Gershenson$^{1}$, Stuart A. Kauffman$^{2,3}$, Ilya Shmulevich$^{4}$ \\
\mbox{}
$^1$Vrije Universiteit Brussel, Krijgskundestraat 33 B-1160 Brussel, Belgium\\
$^2$Institute for Biocomplexity and Informatics, University of Calgary\\
2500 University Drive NW, Calgary, Alberta T2N 1N4, Canada\\
$^3$Santa Fe Institute,1399 Hyde Park Road, Santa Fe, NM, USA 87501\\
$^4$Institute for Systems Biology, 1441 North 34th Street, Seattle, WA, USA 98103-8904 \\
cgershen@vub.ac.be \ skauffman@ucalgary.ca \ is@ieee.org
}

\maketitle

\begin{abstract}
Evolution depends on the possibility of successfully exploring fitness
landscapes via mutation and recombination. With these search procedures,
exploration is difficult in ``rugged" fitness landscapes, where small
mutations can drastically change functionalities in an organism. Random
Boolean networks (RBNs), being general models, can be used to explore
theories of how evolution can take place in rugged landscapes; or even
change the landscapes.

In this paper, we study the effect that redundant nodes have on the
robustness of RBNs. Using computer simulations, we have found that the
addition of redundant nodes to RBNs increases their robustness. We
conjecture that redundancy is a way of ``smoothening" fitness landscapes.
Therefore, redundancy can facilitate evolutionary searches. However, too
much redundancy could reduce the rate of adaptation of an evolutionary
process. Our results also provide supporting evidence in favour of Kauffman's conjecture \cite[p.195]{Kauffman2000}.
\end{abstract}

\section{Introduction}

A system is robust if it continues to function in the face of perturbations 
\cite{Wagner2005}. This makes robustness a desireable property in any system
surrounded by a complex environment. Living systems fall into this category.
Systems able to cope with perturbations will survive and reproduce.
Moreover, robustness plays a key role in evolvability: it allows the gradual
exploration of new solutions while maintaining functionality. A small change
in a fragile system would destroy it. This is not favoured if the
exploration method consists of random mutations.

There are several mechanisms that help a system to be robust \cite%
{vonNeumann1956}, such as modularity \cite{Simon1996,WatsonPollack2005}, degeneracy \cite%
{FernandezSole2003}, distributed robustness \cite{Wagner2004}, and
redundancy. Here, we focus on this last one. It consists on having more than
one copy of an element type. If one fails or changes its function, another
can still perform as expected. Redundancy is widespread in genomes of higher
organisms \cite{NowakEtAl1997}. A clear example is seen with diploidy, where each cell has two complete sets of genes. This has several advantages for evolvability, e.g. it provides robustness to recessive detrimental mutations of one gene of a pair. There is also substantial evidence for the hypothesis that growth in genetic regulatory networks occurs primarily via the duplication of genes and subsequent mutations of one or both members of the duplicate pair \cite{FosterEtAl2005,Lewin2000}.
 Thus, it would be desireable to obtain a
better understanding of its mechanisms. To achieve this, we used random
Boolean networks (RBNs) \cite%
{Kauffman1969,Kauffman1993,Wuensche1998,AldanaEtAl2003,Gershenson2004c}, a
very popular model of genetic regulatory networks.

Our present work was initially motivated by Kauffman's conjecture \cite[p.
195]{Kauffman2000}. This states that 1-bit mutations to a \emph{minimal
program} will change drastically the output of the program, making it
indistinguishable from noise. The conjecture points to the necessity of
having some redundancy to allow smooth transitions as a program changes in
an evolutionary search.

However, it is not obvious to obtain a minimal program in a standard
programming language such as C or Java. Still, we can consider cellular
automata (CA) \cite{vonNeumann1966} or RBNs as programs. It has been shown that
certain CA can perform universal computations \cite{Gardner1983,Cook2004}.
Since RBNs are a well studied generalization of CA \cite{Gershenson2002e},
they seem suitable candidates for studying the effect of redundancy in their
robustness. We can measure the redundancy of a RBN, make 1-bit mutations to
the networks, and then measure their robustness comparing the state spaces
of the mutant and the ``original" networks.

It is not obvious how to measure redundancy, or compressibility. Interpreting results from Kolmogorov, Solomonoff, and Chaitin \cite{Chaitin1975} 
we can say that it is impossible to show that a
program is ``minimal" (independently of a fixed universal model of computation). We decided not to attempt to build ``minimal" networks,
but to compare more and less redundant networks, and try to find differences.

This paper is organized as follows: In the following section, we present
random Boolean networks. Then we introduce a method for adding redundant nodes in RBNs. We used this method in simulations to measure the robustness of RBNs to point mutations.
A discussion of our results follows, to finally present concluding remarks.

\section{Random Boolean networks}

A random Boolean network (RBN) \cite%
{Kauffman1969,Kauffman1993,Wuensche1998,AldanaEtAl2003,Gershenson2004c} has $%
N$ nodes (consisting each of one Boolean variable) $\{\sigma _{1},\sigma
_{2},\cdots ,\sigma _{N}\}$ that can take values of zero or one. The state
of each node $\sigma _{i}$\ is determined by $k_{i}$ randomly chosen nodes
(connections) $\{\sigma _{i_{1}},\cdots ,\sigma _{i_{k_{i}}}\}$. In the
classic model, $k_{i}=k_{j},$ $\forall $ $i,j$, so every node has $K$ inputs.

The way in which the $k_{i}$ elements determine the value of $\sigma _{i}$
is given by a Boolean function $f_{i}(\sigma _{i_{1}},\cdots ,\sigma
_{i_{k_{i}}}).$ Each combination of inputs (there are $2^{k_{i}}$) will
return $f_{i}=1$ with a probability $p$ and $f_{i}=0$ with a probability $%
1-p $. Once the connections and functions have been chosen randomly, the
network structure does not change. In the classical model, the elements are
updated synchronously:

\begin{equation}
\sigma _{i}(t+1)=f_{i}(\sigma _{i_{1}}(t),\cdots ,\sigma _{i_{k_{i}}}(t))
\end{equation}

$f_{i}$ can be represented as a lookup table, where the rightmost column
represents $\sigma _{i}(t+1)$ determined by $k_{i}$ inputs represented in
the rest of the columns exhaustively combined in $2^{k_{i}}$ rows.

Since the number of possible states is finite ($2^{N}$), sooner or later the
network will reach a state that has been already visited. Because the
dynamics are deterministic, the network has reached an \emph{attractor}. If
we use RBNs as models of genetic regulatory networks, attractors can be seen
as cell types \cite{HuangIngber2000}.

The dynamics of ``families" or ensembles \cite{Kauffman2004} of networks can
be studied to find statistical properties common in networks independently
of their precise functionalities. We will use this ensemble approach to
study the effect of redundancy in families of RBNs.

\section{Introducing Redundancy}

At first we attempted to study the effect of redundant links in RBNs.
However, redundant links are fictitious links i.e. the functionality does
not change when they are removed. We began with a network with a desired
percentage of fictitious inputs (this can be regulated with probability $p$
of having 1's in lookup tables), and then measured its stability. Afterwards
we reduced the number of fictitious inputs, measuring the stability until
there were no redundant links. The way we measured stability was: first,
sensitivity to initial conditions, comparing if similar initial states
converged or diverged. Second, making 1-bit mutants of the lookup tables of
the networks, and then comparing the overlap of the state space transitions
using normalised Hamming distances (\ref{hamming}). Our result: there was no
difference. First, since fictitious inputs have no functionality, they do
not affect the convergence of the networks. Second, it seems that the
probability of making a mutation to a redundant link (which did not have
functionality) and making it functional is the same as doing the opposite
(making a functional link non-functional). Therefore, the redundancy of
links does not affect the stability of networks.

\begin{equation}
H(A,B)=\frac{1}{n}\sum\limits_{i}^{n}\left\vert a_{i}-b_{i}\right\vert
\label{hamming}
\end{equation}

We realized that the redundancy in nodes was very different from the
redundancy in links.\footnote{%
This was inspired in redundancy of evolvable hardware \cite{Thompson1998}}
It would be computationally complicated to generate a network with some
redundant nodes, and then trim them. Therefore we did the opposite: RBNs
were generated randomly, and then redundant nodes were added.

The method we developed for adding a redundant node to a RBN is the
following:

\begin{enumerate}
\item Select randomly a node $X$ to be ``duplicated".

\item Add a new node $R$ to the network ($N^{new}=N^{old}+1$), with the same
inputs and lookup table as $X$ (i.e. $k_{R}=k_{X},$ $f_{R}=f_{X}$), and
outputs to the same nodes of which $X$ is input:%
\begin{equation}
k_{i}^{new}=k_{i}^{old}\cup k_{i_{R}}\quad if~\exists k_{i_{X}},\forall i
\end{equation}

\item Double the lookup tables of the nodes of which $X$ is input with the
following criterion: When $R=0$, copy the old lookup table. When $R=1$, and $%
X=0$, copy the same values for all combinations when $X=1$ and $R=0$. Copy
again the same values to the combinations where $X=1$ and $R=1$. In other
words, make an inclusive OR function in which $X$ OR $R$ should be one, to
obtain the old outputs when only $X$ was one%
{\tiny \begin{equation*}
\left. 
\begin{array}{c}
f_{i}^{new}(\sigma _{i_{1}},\cdots ,\sigma _{i_{X}}=0,\sigma
_{i_{R}}=0\cdots ,\sigma _{i_{k_{i}}})=f_{i}^{old}(\sigma _{i_{1}},\cdots
,\sigma _{i_{X}}=0,\cdots ,\sigma _{i_{k_{i}}}) \\ 
f_{i}^{new}(\sigma _{i_{1}},\cdots ,\sigma _{i_{X}}=0,\sigma
_{i_{R}}=1\cdots ,\sigma _{i_{k_{i}}})=f_{i}^{old}(\sigma _{i_{1}},\cdots
,\sigma _{i_{X}}=1,\cdots ,\sigma _{i_{k_{i}}}) \\ 
f_{i}^{new}(\sigma _{i_{1}},\cdots ,\sigma _{i_{X}}=1,\sigma
_{i_{R}}=0\cdots ,\sigma _{i_{k_{i}}})=f_{i}^{old}(\sigma _{i_{1}},\cdots
,\sigma _{i_{X}}=1,\cdots ,\sigma _{i_{k_{i}}}) \\ 
f_{i}^{new}(\sigma _{i_{1}},\cdots ,\sigma _{i_{X}}=1,\sigma
_{i_{R}}=1\cdots ,\sigma _{i_{k_{i}}})=f_{i}^{old}(\sigma _{i_{1}},\cdots
,\sigma _{i_{X}}=1,\cdots ,\sigma _{i_{k_{i}}})%
\end{array}%
\right\} \quad if~\exists k_{i_{X}},\forall i
\end{equation*}
}\end{enumerate}

\begin{figure*}[tbp]
\begin{center}
\includegraphics[
height= 6.8952cm, width=10.8074cm]
{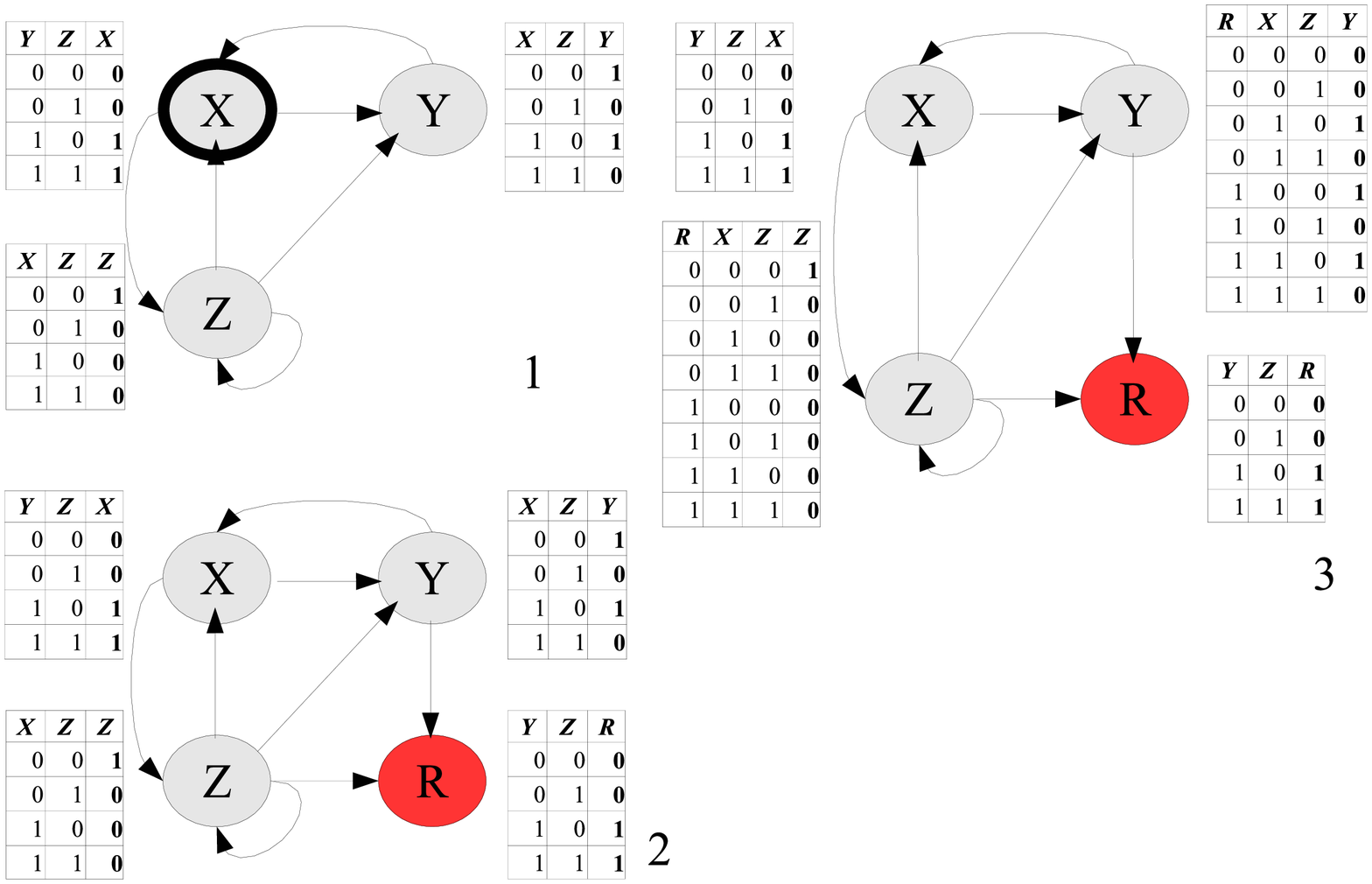}
\end{center}
\caption{Example of addition of a red node to a RBN $N=3$, $K=2$ (see text for details)}
\label{addrednode}
\end{figure*}

After this, $R$ will be a redundant node of $X$, and vice versa. We will
call R, and other redundant nodes which are \emph{added} to the original
network, ``\emph{red}" nodes. The ``original" nodes of the network will be
called ``\emph{white}" nodes. A diagram illustrating the inclusion of a red
node is depicted in Figure \ref{addrednode}. Lookup tables show how the
node of the rightmost column (in bold) at time $t+1$ will be affected by the
values of the other columns at time $t$.

We implemented this algorithm in RBNLab \cite{RBNLab}, and used this
software laboratory to measure the overlaps of state space transitions of
1-bit mutant nets with ``original" ones, as we add more and more redundant
nodes. The mutations consist in flipping a random bit from the lookup table
of a randomly selected node. We used normalised Hamming distances (\ref{hamming}) to measure the difference $dS$ between state spaces:

\begin{equation}
dS = \frac{1}{2^{n}}\sum\limits_{i}^{2^{n}} H(A_{i}^{t+1},M_{i}^{t+1})  \label{dS}
\end{equation}

where $A$ is the ``original" network, and $M$ is a 1-bit mutant of $A$. $dS$ is calculated by computing one time step in both networks  for each initial state $i$. Then, the difference of the states that both networks transitioned is calculated with (\ref{hamming}). This is averaged for all initial states ($2^{N}$), and the result reflects how similar the transitions are in both state spaces. If $dS=0$, there is no difference between state spaces,
and thus the mutation had no effect. A higher $dS$ reflects a greater effect
of the mutation in the state space. There is no correlation of state spaces,
i.e. a mutation is maximally catastrophic, when $dS\simeq 0.5$.

A disadvantage of this method is that it is restricted to small networks,
since the state space is $2^{N}$. Therefore, for each node that is added to
a network, its state space is doubled. Another option would be to have
non-exhaustive explorations of state spaces, but this could be misleading. Therefore, we decided to limit our simulations to small networks.

\section{Simulation Results}

\begin{figure*}[tbp]
\begin{center}
\includegraphics[
height= 3.1116in, width=3.7879in]
{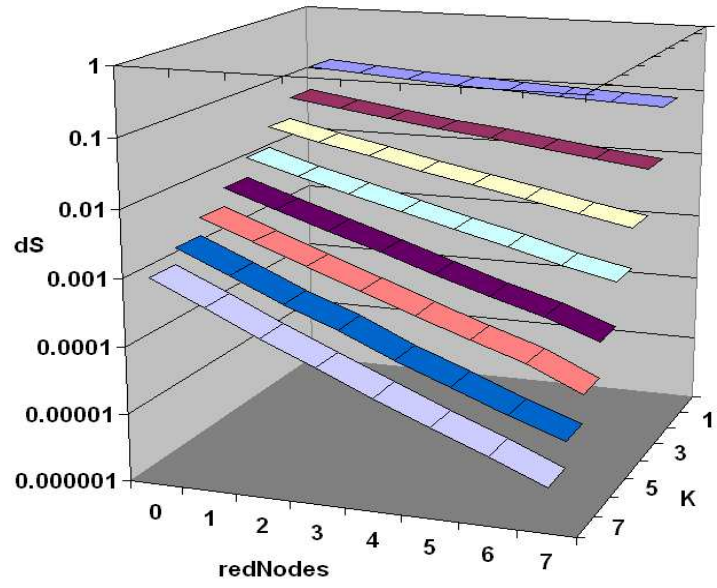}
\end{center}
\caption{$N=7$, average $dS$ for 10000 networks}
\label{n7}
\end{figure*}

We can appreciate the average results of ten thousand networks of $N=7$, $p=0.5$ and
different $K$ values in Figure \ref{n7}. We can see that $dS$ decreases
exponentially as red nodes are added. We should note that the $dS$ for RBNs
without red nodes decreases as $K$ increases. This is because the lookup
tables \ are doubled each time $K$ is incremented. Thus, a one-bit mutation
will have less effect on a network with higher connectivity. To overcome
this, we plotted the values of $dS^{K}$ for the same results in Figure \ref%
{n7dSK}. We can see that a 1-bit mutation makes $dS^{K}\simeq 0.1$ for networks without red nodes. Clearly
the addition of red nodes increases the robustness of the networks. The
effect of red nodes is more evident for higher $K$ values, where the network
dynamics are more chaotic \cite{Gershenson2004c}.

\begin{figure*}[tbp]
\begin{center}
\includegraphics[
height= 3.1116in, width=3.7879in]
{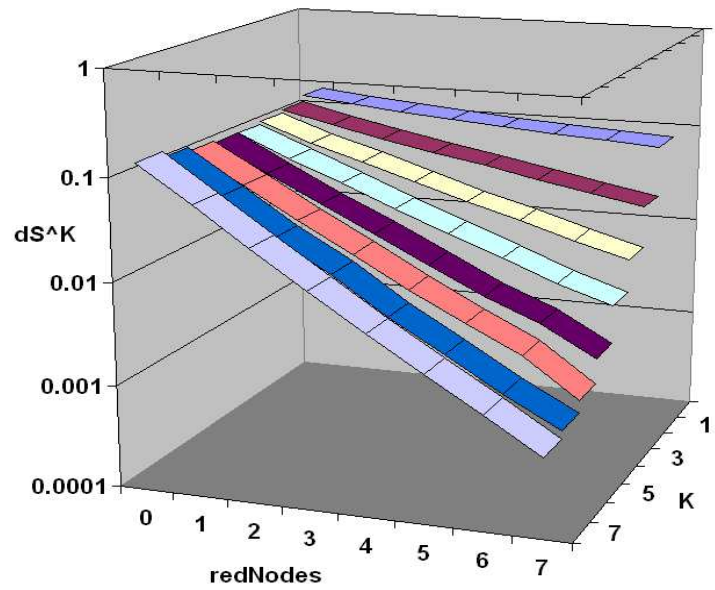}
\end{center}
\caption{$N=7$, average $dS^{K}$ for 10000 nets}
\label{n7dSK}
\end{figure*}

\begin{figure*}[tbp]
\begin{center}
\includegraphics[
height= 3.1116in, width=3.7879in]
{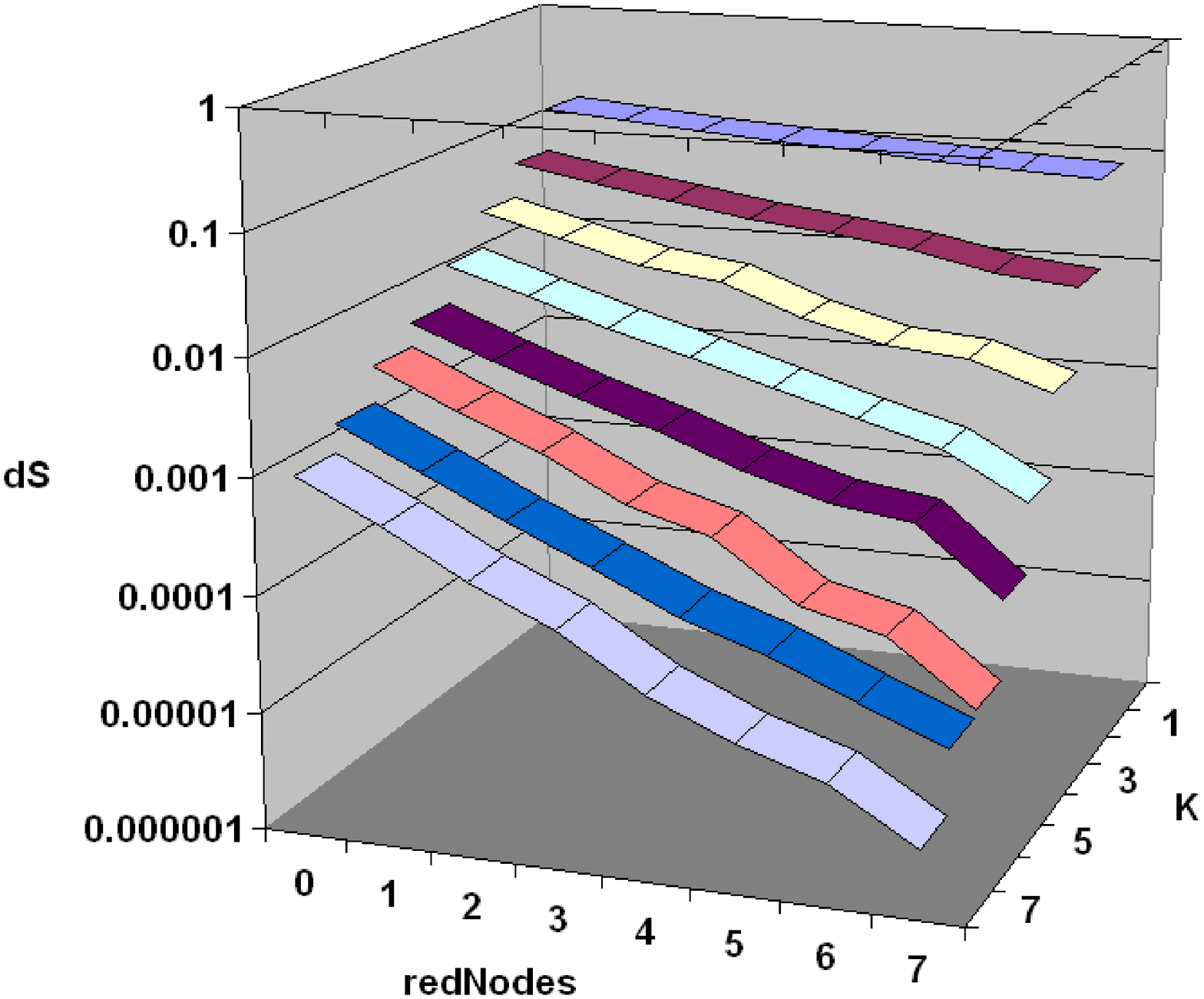}
\end{center}
\caption{$N=10$, average $dS$ for 1000 networks}
\label{n10}
\end{figure*}

To be certain that the robustness is given by the addition of a red node,
and not of any node, we performed simulations with $N=10$. The average
values of $dS$ for one thousand networks are shown in Figure \ref{n10}. We
can quickly see that a network with seven white nodes and three red nodes
has a lower difference in state spaces than a network with ten white nodes
and no red nodes, especially for high values of $K$.

We obtained results of RBNs with scale free topology \cite{Aldana2003} (data
not shown). We observed a similar tendency: as red nodes are introduced in a
network, mutants will have less differences in the state space, especially
for chaotic networks of high connectivity.

We also obtained results for different updating schemes \cite%
{HarveyBossomaier1997,Gershenson2002e,GershensonEtAl2003a} and both
topologies, with similar results (data not shown). The precise effect of
redundant nodes on $dS$ does change with the updating scheme: in all
cases $dS$ decreases as redundant nodes are added.

\section{Discussion}

It is intuitive to assume that redundancy will increase the robustness of a
system. If in a minimal program all nodes matter, any mutation will have
more drastic effects in the output than in a non-optimal program. This has
been seen experimentally in evolvable hardware \cite{Thompson1998}, where
some redundancy allows better evolution of circuits, but too much redundany
obstructs it. Still, the goal of our work was to quantify the effect of node
redundancy in robustness to mutations in RBN.

Fictitious inputs do not increase the robustness because mutating such
inputs can change the logic function. This is not the case for redundant
nodes: if a node is mutated, the rest of the network keeps its (partial)
function. In other words, mutations in redundant nodes do not propagate
through the network, whereas mutations in fictitious inputs do.

From our results we can say that redundancy $smoothens$ rough fitness
landscapes \cite{Kauffman1993}, making evolutionary search potentially feasible in rough fitness landscapes.
However, too much redundancy should be detrimental for evolvability, because
evolutionary search becomes slower. We can also confirm recent results by
Andreas Wagner \cite{Wagner2005b} and notice that red nodes increase the \emph{neutrality} 
\cite{Kimura1983} of the network. A neutral mutation is one that has no
effect in the function (or phenotype). As we can see, redundancy of nodes
decreases the probability of a mutation causing a change in the
functionality. If a mutation occurs in a red node, another node will take
its role to perform the old function. From this we can suggest people working
with genetic algorithms: if a fitness landscape is too rough to make an
evolutionary search on it, then add some redundancy that will smoothen it. Nevertheless, not all types of redundancy will be useful \cite{HarveyThompson1997}.

It could be argued that ``useless" redundancy can be added to a state space, i.e. with no effect on the function, and thus would be of no use for evolvability. However, in our RBN approach, a red node cannot be useless, since it is connected to the rest of the network and thus affects its behaviour. Therefore, any red node will have a potential effect on the function of a \emph{mutated} network: the function of the mutated network will be close to that of the original network, for another node is duplicating the functionality that changed with the mutation. In other words, red nodes provide ``useful" redundancy. On the other hand, redundancy given by fictitious links in lookup tables is indeed ``useless": adding fictitious links will have no effect on the mutated network functionality. Still, a general mechanism for deciding wether a certain type of redundancy is useful or not seems to be an open question. In our case, node redundancy does help an evolutionary search in rugged fitness landscapes. To which cases this is applicable remains to be explored. For example, if a fitness landscape is already smooth, adding redundant nodes will be useless to an evolutionary search.

Returning to the original motivation of this work, if Kauffman's conjecture holds, a random mutation on a ``minimal" RBN should
produce a $dS\simeq 0.5$. However, we cannot prove if a RBN cannot be
reduced or compressed independently of a fixed universal model of computation. Thus, we cannot know if a RBN is minimal. What we
have seen is that ``less compressible" RBNs, i.e. with no red nodes, have
always a higher $dS$. As we add more red nodes, the RBN becomes more
compressible, and $dS$ decreases, showing an increase in robustness. As we
have seen, the compressibility is directly proportional to the robustness of
a network to random mutations. We could extrapolate and induce that if the
compressibility could be minimal, then the robustness would be also minimal.
The minimal robustness would be given when a single mutation would create $%
dS\simeq 0.5$, so we can conclude that this corresponds to a minimal
compressibility, i.e. a minimal program, even when in theory it is not
possible to show this. This does not prove Kauffman's conjecture, but
provides reasonable evidence for its validity.

\section{Conclusions}

We have presented an algorithm for introducing redundant nodes in a random
Boolean networks. We used this to show that redundancy \emph{of nodes} increases
the mutational robustness of RBNs. This may have advantages for evolvability, depending on the particular problem domain.
However, there are other issues that should be also considered for
evolvability, such as distributed robustness, degeneracy,
and modularity. The relationship between neutrality
and redundancy demands further study. This is also the case for redundancy
of nodes in larger networks.

\section{Ackwnoledgements}

We should like to thank Andrew Wuensche, Adrian Thompson, Miguel Garvie, Tatiana Kalganova, Cosma Shalizi, and Inman Harvey for interesting discussions and comments. C.G. was partly supported by the Consejo Nacional de
Ciencia y Teconolg\'{\i}a (CONACyT) of M\'{e}xico.

\bibliography{carlos,RBN,sos}

\begin{thebibliography}{}

\bibitem[Aldana, 2003]{Aldana2003}
Aldana, M. (2003).
\newblock Boolean dynamics of networks with scale-free topology.
\newblock {\em Physica D}, 185(1):45--66.

\bibitem[Aldana-Gonz{\'a}lez et~al., 2003]{AldanaEtAl2003}
Aldana-Gonz{\'a}lez, M., Coppersmith, S., and Kadanoff, L.~P. (2003).
\newblock Boolean dynamics with random couplings.
\newblock In Kaplan, E., Marsden, J.~E., and Sreenivasan, K.~R., editors, {\em
  Perspectives and Problems in Nonlinear Science. A Celebratory Volume in Honor
  of Lawrence Sirovich}. Springer Applied Mathematical Sciences Series.

\bibitem[Chaitin, 1975]{Chaitin1975}
Chaitin, G.~J. (1975).
\newblock Randomness and mathematical proof.
\newblock {\em Scientific American}, 232(5):47--52.

\bibitem[Cook, 2004]{Cook2004}
Cook, M. (2004).
\newblock Universality in elementary cellular automata.
\newblock {\em Complex Systems}, 15(1):1--40.

\bibitem[Fern{\'a}ndez and Sol{\'e}, 2004]{FernandezSole2003}
Fern{\'a}ndez, P. and Sol{\'e}, R. (2004).
\newblock The role of computation in complex regulatory networks.
\newblock In Koonin, E.~V., Wolf, Y.~I., and Karev, G.~P., editors, {\em Power
  Laws, Scale-Free Networks and Genome Biology}. Landes Bioscience.

\bibitem[Foster et~al., 2005]{FosterEtAl2005}
Foster, D.~V., Kauffman, S.~A., and Socolar, J. E.~S. (2005).
\newblock Network growth models and genetic regulatory networks.
\newblock arXiv q-bio.MN/0510009.

\bibitem[Gardner, 1983]{Gardner1983}
Gardner, M. (1983).
\newblock {\em Wheels, Life, and Other Mathematical Amusements}, chapter 20-22.
\newblock W. H. Freeman.

\bibitem[Gershenson, 2002]{Gershenson2002e}
Gershenson, C. (2002).
\newblock Classification of random {Boolean} networks.
\newblock In Standish, R.~K., Bedau, M.~A., and Abbass, H.~A., editors, {\em
  Artificial Life {VIII}: Proceedings of the Eight International Conference on
  Artificial Life}, pages 1--8. MIT Press.

\bibitem[Gershenson, 2004]{Gershenson2004c}
Gershenson, C. (2004).
\newblock Introduction to random boolean networks.
\newblock In Bedau, M., Husbands, P., Hutton, T., Kumar, S., and Suzuki, H.,
  editors, {\em Workshop and Tutorial Proceedings, Ninth International
  Conference on the Simulation and Synthesis of Living Systems {(ALife} {IX)}},
  pages 160--173, Boston, MA.

\bibitem[Gershenson, 2005]{RBNLab}
Gershenson, C. (2005).
\newblock {RBNLab}.
\newblock http://rbn.sourceforge.net.

\bibitem[Gershenson et~al., 2003]{GershensonEtAl2003a}
Gershenson, C., Broekaert, J., and Aerts, D. (2003).
\newblock Contextual random {Boolean} networks.
\newblock In Banzhaf, W., Christaller, T., Dittrich, P., Kim, J.~T., and
  Ziegler, J., editors, {\em Advances in Artificial Life, 7th European
  Conference, {ECAL} 2003 {LNAI} 2801}, pages 615--624. Springer-Verlag.

\bibitem[Harvey and Bossomaier, 1997]{HarveyBossomaier1997}
Harvey, I. and Bossomaier, T. (1997).
\newblock Time out of joint: Attractors in asynchronous random {Boolean}
  networks.
\newblock In Husbands, P. and Harvey, I., editors, {\em Proceedings of the
  Fourth European Conference on Artificial Life {(ECAL97)}}, pages 67--75. MIT
  Press.

\bibitem[Harvey and Thompson, 1997]{HarveyThompson1997}
Harvey, I. and Thompson, A. (1997).
\newblock Through the labyrinth evolution finds a way: A silicon ridge.
\newblock In Higuchi, T., Iwata, M., and Weixin, L., editors, {\em Proceedings
  of the First International Conference on Evolvable Systems: From Biology to
  Hardware}, volume 1259 of {\em LNCS}, pages 406 -- 422. Springer-Verlag.

\bibitem[Huang and Ingber, 2000]{HuangIngber2000}
Huang, S. and Ingber, D.~E. (2000).
\newblock Shape-dependent control of cell growth, differentiation, and
  apoptosis: Switching between attractors in cell regulatory networks.
\newblock {\em Exp. Cell Res.}, 261:91--103.

\bibitem[Kauffman, 1969]{Kauffman1969}
Kauffman, S.~A. (1969).
\newblock Metabolic stability and epigenesis in randomly constructed genetic
  nets.
\newblock {\em Journal of Theoretical Biology}, 22:437--467.

\bibitem[Kauffman, 1993]{Kauffman1993}
Kauffman, S.~A. (1993).
\newblock {\em The Origins of Order}.
\newblock Oxford University Press.

\bibitem[Kauffman, 2000]{Kauffman2000}
Kauffman, S.~A. (2000).
\newblock {\em Investigations}.
\newblock Oxford University Press.

\bibitem[Kauffman, 2004]{Kauffman2004}
Kauffman, S.~A. (2004).
\newblock The ensemble approach to understand genetic regulatory networks.
\newblock {\em Physica A: Statistical Mechanics and its Applications},
  340(4):733--740.

\bibitem[Kimura, 1983]{Kimura1983}
Kimura, M. (1983).
\newblock {\em The Neutral Theory of Molecular Evolution}.
\newblock Cambridge University Press, Cambridge.

\bibitem[Lewin, 2000]{Lewin2000}
Lewin, B. (2000).
\newblock {\em Genes VII}.
\newblock Oxford University Press.

\bibitem[Nowak et~al., 1997]{NowakEtAl1997}
Nowak, M.~A., Boerlijst, M.~C., Cooke, J., and {Maynard Smith}, J. (1997).
\newblock Evolution of genetic redundancy.
\newblock {\em Nature}, 388:167--171.

\bibitem[Simon, 1996]{Simon1996}
Simon, H.~A. (1996).
\newblock {\em The Sciences of the Artificial}.
\newblock MIT Press, 3rd edition.

\bibitem[Thompson, 1998]{Thompson1998}
Thompson, A. (1998).
\newblock {\em Hardware Evolution: Automatic Design of Electronic Circuits in
  Reconfigurable Hardware by Artificial Evolution}.
\newblock Distinguished dissertation series. Springer-Verlag.

\bibitem[{von Neumann}, 1956]{vonNeumann1956}
{von Neumann}, J. (1956).
\newblock Probabilistic logics and the synthesis of reliable organisms from
  unreliable components.
\newblock In Shannon, C. and McCarthy, J., editors, {\em Automata Studies},
  Princeton. Princeton University Press.

\bibitem[{von Neumann}, 1966]{vonNeumann1966}
{von Neumann}, J. (1966).
\newblock {\em The Theory of Self-Reproducing Automata}.
\newblock University of Illinois Press.
\newblock Edited by A. W. Burks.

\bibitem[Wagner, 2004]{Wagner2004}
Wagner, A. (2004).
\newblock Distributed robustness versus redundancy as causes of mutational
  robustness.
\newblock Technical Report 04-06-018, Santa Fe Institute.

\bibitem[Wagner, 2005a]{Wagner2005}
Wagner, A. (2005a).
\newblock {\em Robustness and Evolvability in Living Systems}.
\newblock Princeton University Press, Princeton, NJ.

\bibitem[Wagner, 2005b]{Wagner2005b}
Wagner, A. (2005b).
\newblock Robustness, neutrality, and evolvability.
\newblock {\em FEBS Letters}, 579:1772--1778.

\bibitem[Watson and Pollack, 2005]{WatsonPollack2005}
Watson, R.~A. and Pollack, J.~A. (2005).
\newblock Modular interdependency in complex dynamical systems.
\newblock {\em Artificial Life}, 11(4):445--457.

\bibitem[Wuensche, 1998]{Wuensche1998}
Wuensche, A. (1998).
\newblock Discrete dynamical networks and their attractor basins.
\newblock In Standish, R., Henry, B., Watt, S., Marks, R., Stocker, R., Green,
  D., Keen, S., and Bossomaier, T., editors, {\em Complex Systems '98}, pages
  3--21, University of New South Wales, Sydney, Australia.

\end{thebibliography}
\bibliographystyle{alife10}

\end{document}